\documentclass[aps,superscriptaddress,twocolumn]{revtex4}
\usepackage{amsmath}
\usepackage{graphicx}
\usepackage{dcolumn}
\usepackage{verbatim}
\usepackage{times}
\usepackage{subfigure}
\usepackage{bm}
\usepackage{color}
\usepackage[colorlinks,citecolor=blue,linkcolor=blue,dvipdfm]{hyperref}
\usepackage{subfigure}

\usepackage{booktabs}
\usepackage{appendix}

\begin{document}

\title{Effective model and electron correlations in trilayer nickelate superconductor La$_4$Ni$_3$O$_{10}$}

\author{Peng-Fei Tian}
\affiliation{College of Physics and Electronic Information Engineering, Guilin University of Technology, Guilin 541004, China}
\author{Hao-Tian Ma}
\affiliation{College of Physics and Electronic Information Engineering, Guilin University of Technology, Guilin 541004, China}
\author{Xing Ming}
\affiliation{College of Physics and Electronic Information Engineering, Guilin University of Technology, Guilin 541004, China}
\author{Xiao-Jun Zheng}
\affiliation{College of Physics and Electronic Information Engineering, Guilin University of Technology, Guilin 541004, China}
\author{Huan Li}
\email{lihuan@glut.edu.cn}
\affiliation{College of Physics and Electronic Information Engineering, Guilin University of Technology, Guilin 541004, China}

\date{\today}

\begin{abstract}

Recently, signatures of superconductivity with critical temperature from 20 to 30 K have been reported in pressured trilayer nickelate La$_4$Ni$_3$O$_{10}$ through a pressure-induced structure transition. Here we explore the evolution of electronic structures and electronic correlations in different phases of La$_4$Ni$_3$O$_{10}$ under corresponding pressure regions, by using density functional theory (DFT) combined with dynamical mean-field theory (DMFT). Similar to bilayer superconductor La$_3$Ni$_2$O$_{7}$, the electronic bands in superconducting La$_4$Ni$_3$O$_{10}$ are dominated by Ni-3$d_{x^2-y^2}$ and 3$d_{z^2}$ orbits near the Fermi level, in contrast, the inner Ni-O plane in La$_4$Ni$_3$O$_{10}$ generates a doublet hole-pocket Fermi surfaces around the Brillouin-zone corner, meanwhile one branch of the Ni-$3d_{z^2}$ bands is pushed very close above the Fermi level, which can induce an electron pocket through small electron doping. The DFT+DMFT simulations suggest that the electronic correlations only give minor modification to the Fermi surfaces, meanwhile the Ni-$3d_{z^2}$ and 3$d_{x^2-y^2}$ states on outer Ni-O layers have considerable greater mass enhancements than on the inner layer. The sensitiveness of electronic structure under doping and unique layer dependence of correlation suggest a distinct superconducting mechanism with respect to bilayer La$_3$Ni$_2$O$_{7}$. Based on the DFT and DFT+DMFT simulations, we eventually derive a trilayer effective tight-binding model, which can produce rather precise electronic bands and Fermi surfaces, hence can serve as an appropriate model to further study the superconducting mechanism and paring symmetry in trilayer La$_4$Ni$_3$O$_{10}$.

\end{abstract}

\maketitle

\section{Introduction}

Recently, experimental evidences have shown that the nickelate La$_3$Ni$_2$O$_{7}$ exhibits high-temperature superconductivity with transition temperature $T_c$ exceeding 80 K under high pressure~\cite{Sun23,Wang23}, which has aroused widespread research interest. Similar to cuprate high-temperature superconductors, the superconductivity of La$_3$Ni$_2$O$_{7}$ arises on bilayer Ni-O planes and can be described by an effective tight-binding model involving $3d_{z^2}$ and $3d_{x^2-y^2}$ orbits of Ni~\cite{YFYang23,Luo23,Gu23}. Although there are still controversies over the superconducting (SC) pairing mechanism and symmetry in La$_3$Ni$_2$O$_{7}$~\cite{Tian23,Qu23,QGYang23,Wang24,Chen23,HQLiu23,Schlomer23}, growing experimental and theoretical evidences suggest the unconventional nature of superconductivity~\cite{Sakakibara23,Shen23,ZLiu23,Hou23,Lu23,JGYang23,Lu2308} and that the driving force of SC pairing is closely related to the magnetic correlations between Ni-$d_{z^2}$ orbits~\cite{Qu23,Shilenko23,Lechermann23,Ouyang23,Liao23,YZhang23,Oh23,Xie24}. Very recently, superconductivity has also been discovered in La$_4$Ni$_3$O$_{10}$ with trilayer Ni-O planes~\cite{Zhang2311,Zhu23}. Likewise, La$_4$Ni$_3$O$_{10}$ also undergoes a structure transition to a higher-symmetric phase before superconductivity emerges~\cite{Li23}. However, unlike the cuprate superconductors with highest $T_c$ in trilayer systems~\cite{Chakravarty04}, the SC transition temperature in trilayer La$_4$Ni$_3$O$_{10}$ ($T_c$ from 20-30 K) is significantly lower than bilayer La$_3$Ni$_2$O$_{7}$~\cite{Sun23}, which poses a new challenge to the theoretical understanding of nickelate superconductivity.

\begin{figure}[tbp]
\hspace{0cm} \includegraphics[totalheight=2.4in]{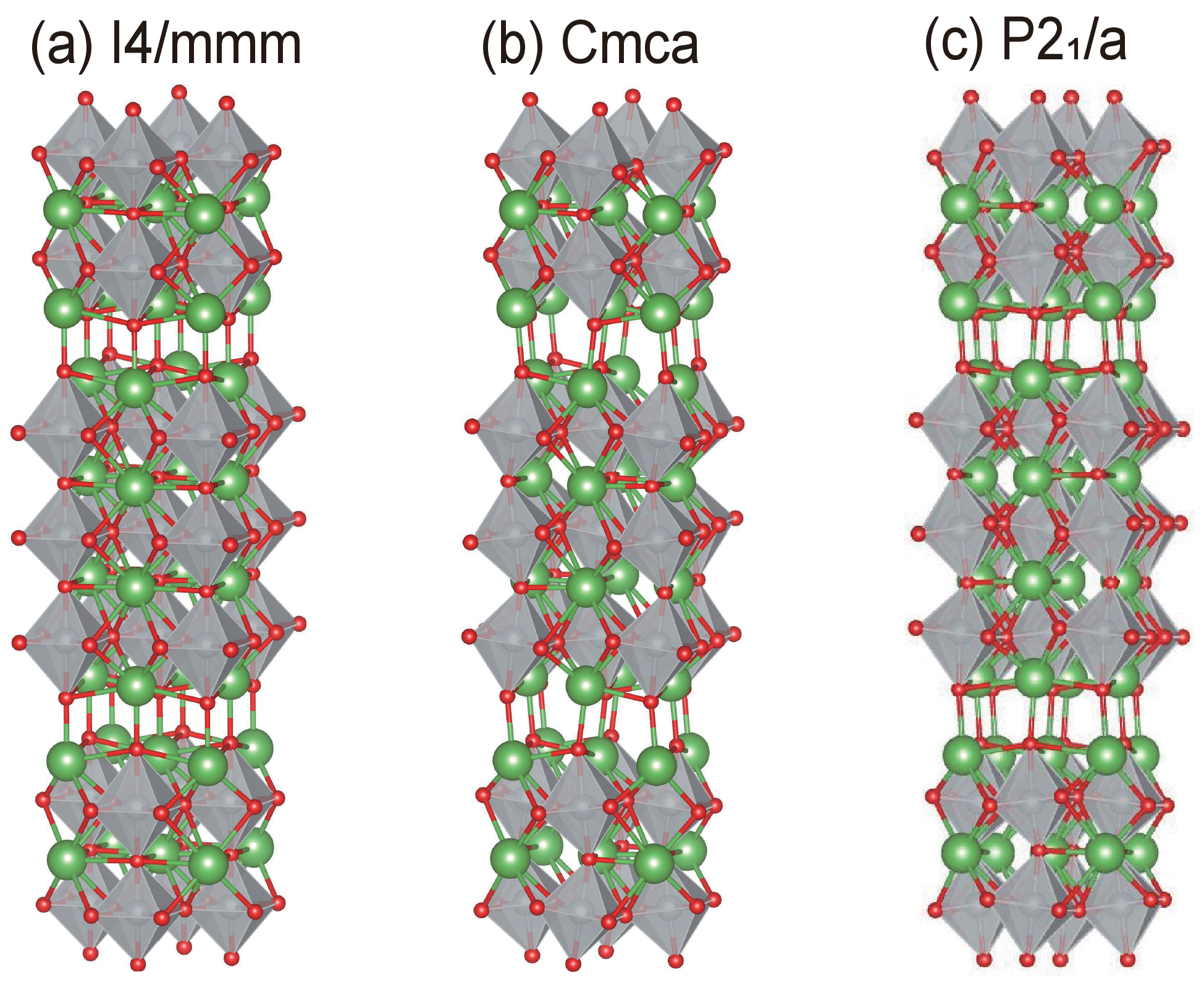}
\caption{Crystal structures of (a) $I4/mmm$, (b) $Cmca$ and (c) $P2_1/a$ phases of La$_4$Ni$_3$O$_{10}$.
}
\label{lattice}
\end{figure}
\begin{figure*}[tbp]
\hspace{0cm} \includegraphics[totalheight=3.4in]{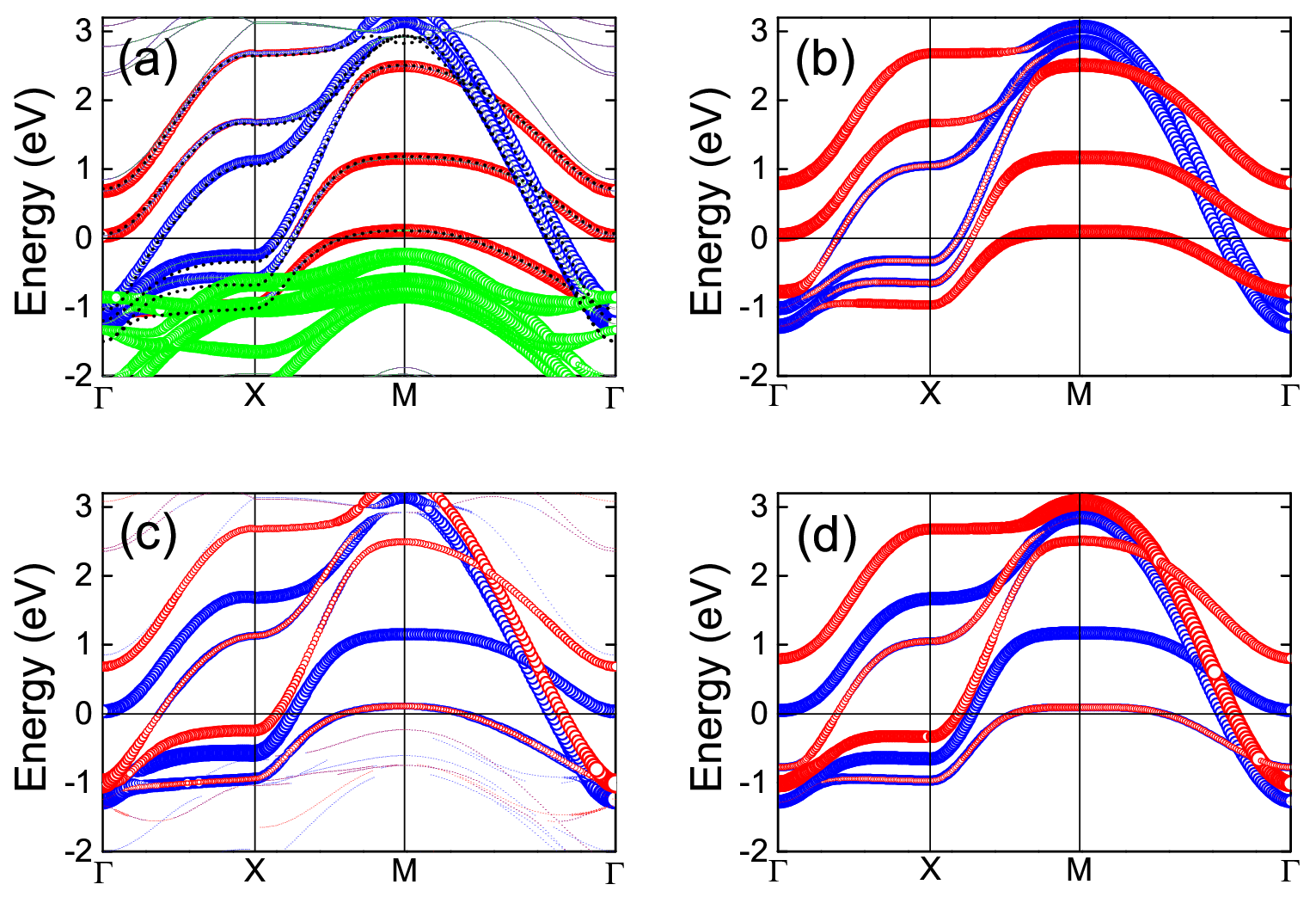}
\caption{ DFT ((a), (c)) and effective-model ((b), (d)) bands for $I4/mmm$ phase of La$_4$Ni$_3$O$_{10}$. In (a), the dashed black lines denote the fitted Wannier bands.
(a) and (b) illustrate the orbital-projected bands, in which the red, blue and green colors denote Ni-$3d_{z^2}$, $3d_{x^2-y^2}$ and $t_{2g}$ bands, respectively. In (c) and (d), red and blue colors denote contributions from inner and outer Ni-O planes, respectively.
}
\label{bands1}
\end{figure*}
In this article, we explore La$_4$Ni$_ 3$O$_{10}$ through first-principle calculations, and reveal the electronic structures of different crystal structures emerging in corresponding pressure region~\cite{Li23,Li17}. We find that the $I4/mmm$ symmetry of the SC phase at high pressure possess a similar electronic structure to the SC phase of La$_3$Ni$_2$O$_{7}$, however, due to the trilayer Ni-O planes, the Fermi surfaces have doublet hole pockets around the Brillouin zone corners, and slight electron doping can give rise to an additional electronic pocket around the Brillouin zone center. Besides, DFT+DMFT calculations illustrate that the electronic correlations is considerably stronger on the outer Ni-O layers than on the inner layer.
In addition, the Fermi surfaces of $Cmca$~\cite{Li17} and $P2_1/a$~\cite{Li23} structures at ambient or low pressures are similar to each other, and can be properly reconstructed by the Fermi surface of $I4/mmm$ phase through Brillouin-zone folding. In this context, we have fitted a tight-binding effective model for SC La$_4$Ni$_ 3$O$_{10}$, which can well describe the electronic and Fermi surface structure thus can serve as an appropriate model for further studying the SC mechanism and pairing symmetry of trilayer nicklate superconductors.

\section{electronic structure and Fermi surfaces}

\begin{figure*}[tbp]
\hspace{0cm} \includegraphics[totalheight=2.5in]{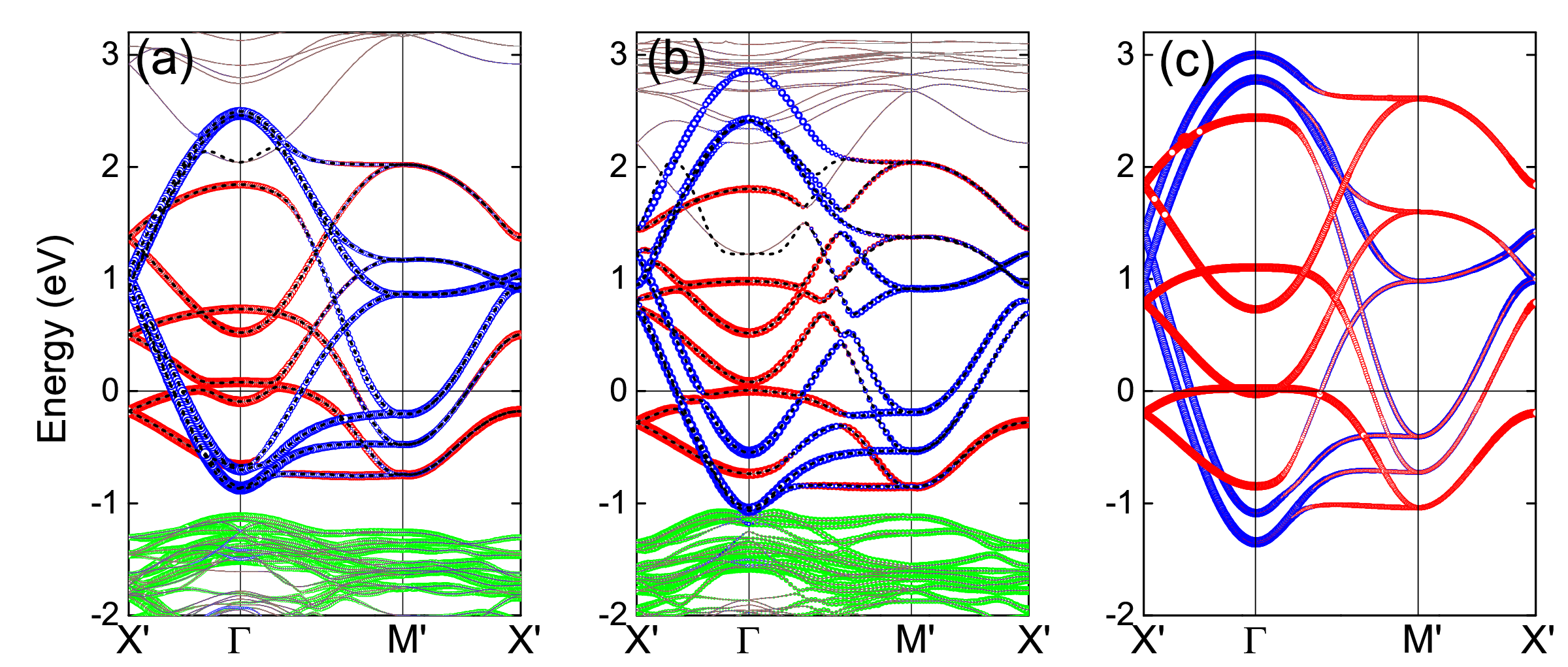}
\caption{
 Projected DFT bands of (a) $Cmca$ and (b) $P2_1/a$ phases, in which the red, blue and green colors denote Ni-$3d_{z^2}$, $3d_{x^2-y^2}$ and $t_{2g}$ components, respectively. (c) Folded bands from the effective model of $I4/mmm$ phase, which can nicely construct the $Cmca$ bands in (a). In (a) and (b), the dashed black lines denote the fitted Wannier bands.
}
\label{bands2}
\end{figure*}

\begin{figure}[tbp]
\hspace{-0cm} \includegraphics[totalheight=2.5in]{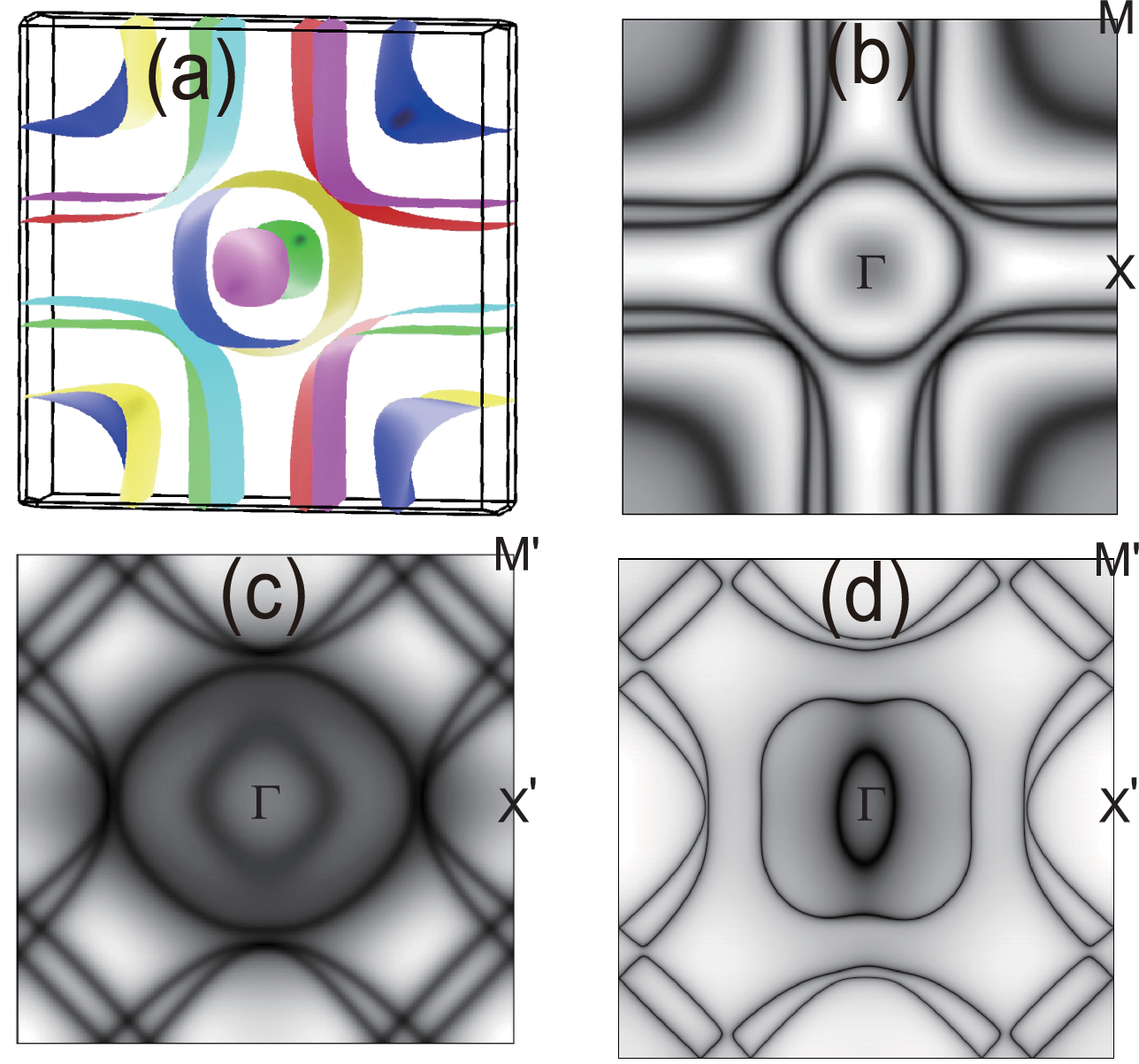}
\caption{Fermi surfaces (FSs) of $I4/mmm$ phase in (a) three dimensional lattice Brillouin zone and (b) on $k_z=0$ plane.
(c) and (d) show the FSs of $Cmca$ and $P2_1/a$ phases on $k_z=0$ plane (corresponds to the folded Brillouin zone in (b)), respectively.
}
\label{FS-wannier}
\end{figure}

Firstly, based on the density functional theory (DFT) embodied in the Vienna ab-initio simulation package (VASP), we discuss the electronic structures of different phases of La$_4$Ni$_3$O$_{10}$ emerging in corresponding pressure regions. Under ambient and low pressures, literatures have reported two similar structures, with $Cmca$~\cite{Li17} or $P2_1/a$~\cite{Li23,Zhang20,MXZhang23} space groups, respectively. At around 12.6 to 13.4 GPa, a structure transition takes place from monoclinic $P2_1/a$ to the tetragonal $I4/mmm$ space group~\cite{Li23}. These three structures are illustrated in Fig. \ref{lattice} (a-c). The orientation of NiO$_6$ octahedrons in $I4/mmm$ is uniform, forming trilayer Ni-O planes with an intra-plane square-lattice structure. While in $Cmca$ and $P2_1/a$ structures, the nearby NiO$_6$ octahedrons are tilted with opposite angles along lattice $\mathbf{a}$ axis (which is along intra-plane next-nearest-neighbor Ni-Ni direction), arranging in a staggered manner along both intra-plane and vertical directions. For $P2_1/a$ phase, the tilted NiO$_6$ octahedrons are further distorted slightly along the Ni-O bonds which are originally perpendicular to the Ni-O planes. For both $Cmca$ and $P2_1/a$ phases, the unit cell on Ni-O planes are doubled than in $I4/mmm$ phase.

\begin{figure}[tbp]
\hspace{-0cm} \includegraphics[totalheight=1.6in]{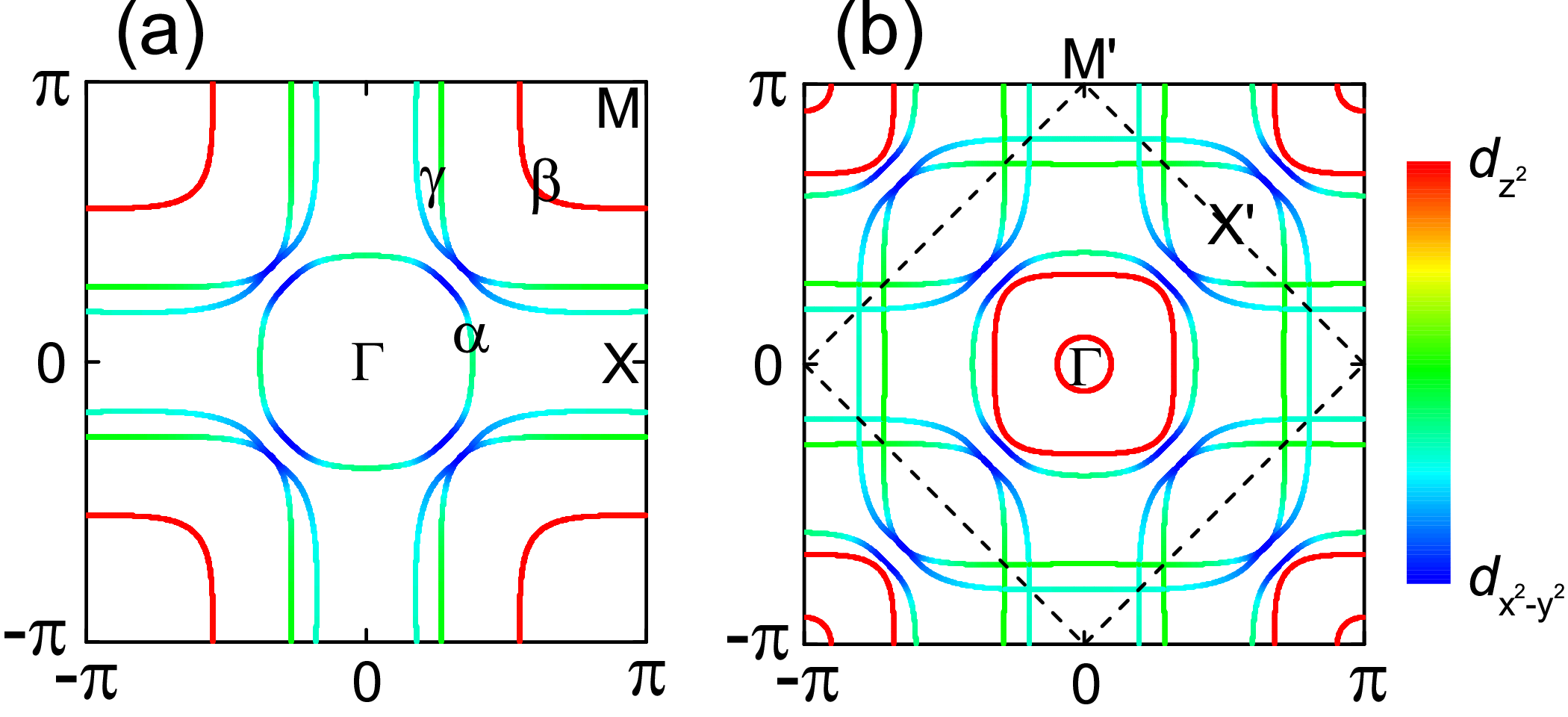}
\caption{(a) FSs of $I4/mmm$ structure from effective two-orbital model. (b) FSs via folding into the contracted Brillouin zone (within the dashed square). In (b), the Fermi energy is set slightly higher than (a), thus additional small electron pockets emerge at the center and corners.
}
\label{FS}
\end{figure}

Fig. \ref{bands1} (a) illustrates the orbital-projected representation of the DFT bands for $I4/mmm$ phase along the high symmetric path on the $k_z=0$ plane in the Brillouin zone, which corresponds to the two-dimensional square-lattice Brillouin zone of Ni-O planes. It can be seen that the electronic bands near the Fermi energy are mainly constructed by $3d_{z^2}$ (red) and $3d_{x^2-y^2}$ (blue) orbits of Ni atoms. Therefore, the Fermi surfaces (FSs) are also composed of these two states, as shown in Fig. \ref{FS-wannier} (a) and (b). The hole pockets around $M$ points at the Brillouin corner are contributed by the $3d_{z^2}$ orbit, while the electron pocket around $\Gamma$ point at the center and the large hole pockets around $M$ are formed by mixture of these two orbits. Fig. \ref{bands1} (c) shows the electronic bands projected by different Ni-O layers. In comparison to La$_3$Ni$_2$O$_{7}$, the inner Ni-O layer in La$_4$Ni$_3$O$_{10}$ gives rise to two additional energy bands (the third and sixth bands from bottom to top, denoted by red), resulting in two nearby hole pockets surrounding $M$ (mixed by two $e_g$ orbits). Besides, the inter-layer interactions through the inner Ni-O layer push the fifth bands (contributed by outer planes) much closer to the Fermi level, hence induces a small electron pocket (by $3d_{z^2}$ orbit) near $\Gamma$ on $k_z\neq0$ plane, see Fig. \ref{FS-wannier}(a).

In $Cmca$ and $P2_1/a$ phases, the NiO$_6$ octahedrons on trilayer Ni-O planes undergo distortion in a staggered manner~\cite{Li17,Li23}, dividing the Ni atoms into two different sublattices. Since the six electronic bands near the Fermi level in $I4/mmm$ phase are mainly constructed by $e_g$ orbits of Ni, therefore, twelve $e_g$ bands emerge near the Fermi level in $Cmca$ and $P2_1/a$ phases, which can be understood in terms of the Brillouin-zone folding of the six $e_g$ bands of $I4/mmm$ phase. The projected electronic bands of $Cmca$ and $P2_1/a$ phases are shown in Fig. \ref{bands2} (a) and (b), respectively. It can also be seen that the twelve Ni-$e_g$ bands in $Cmca$ phase are quite similar to the folded bands (Fig. \ref{bands2} (c)) computed by generalizing the effective model of $I4/mmm$ phase into two-sublattice structure, which can be ascribed to the small deflection angles of the NiO$_6$ octahedrons in $Cmca$ phase. By contrast, the electronic bands of $P2_1/a$ phase are a bit more deviated from the folded bands of $I4/mmm$ phase, due to further tilting of NiO$_6$ octahedrons in $P2_1/a$ phase. As displayed in Fig. \ref{FS-wannier} (c) and (d), the FSs of $Cmca$ and $P2_1/a$ phases are similar to each other and coincide with those reported in Ref.\cite{Zhang2311,Li23}. As will be shown in the following, the FSs of both $Cmca$ and $P2_1/a$ phases can be qualitatively regard as the Brillouin-zone folding of the $I4/mmm$ FSs in Fig. \ref{FS-wannier} (b).

\section{effective tight-binding model}

\begin{figure}[tbp]
\hspace{-0cm} \includegraphics[totalheight=2in]{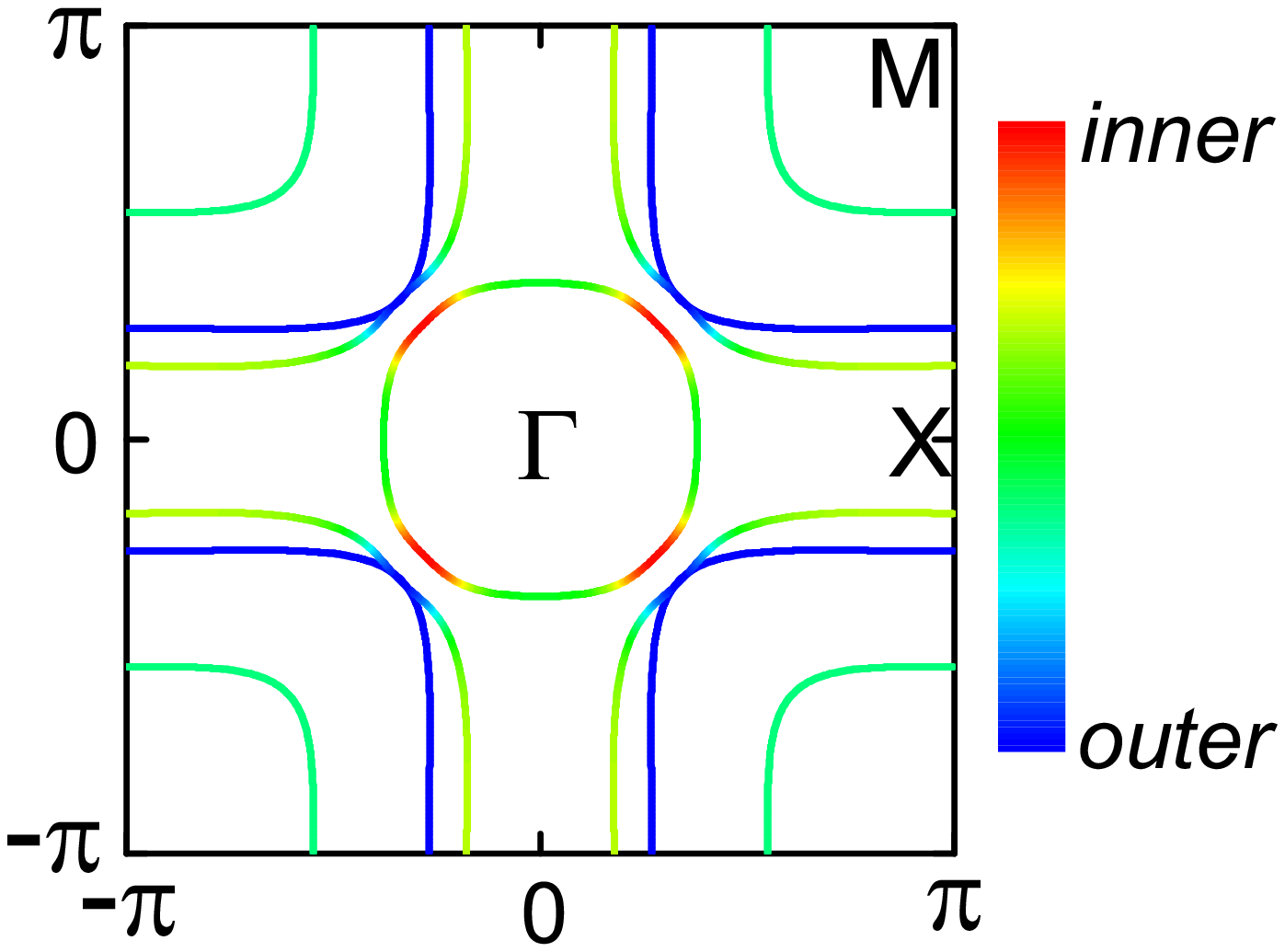}
\caption{Layer contributions of the FSs of $I4/mmm$ structure from effective two-orbital model.
}
\label{FS_layer}
\end{figure}

\begin{figure}[tbp]
\hspace{-0cm} \includegraphics[totalheight=1.6in]{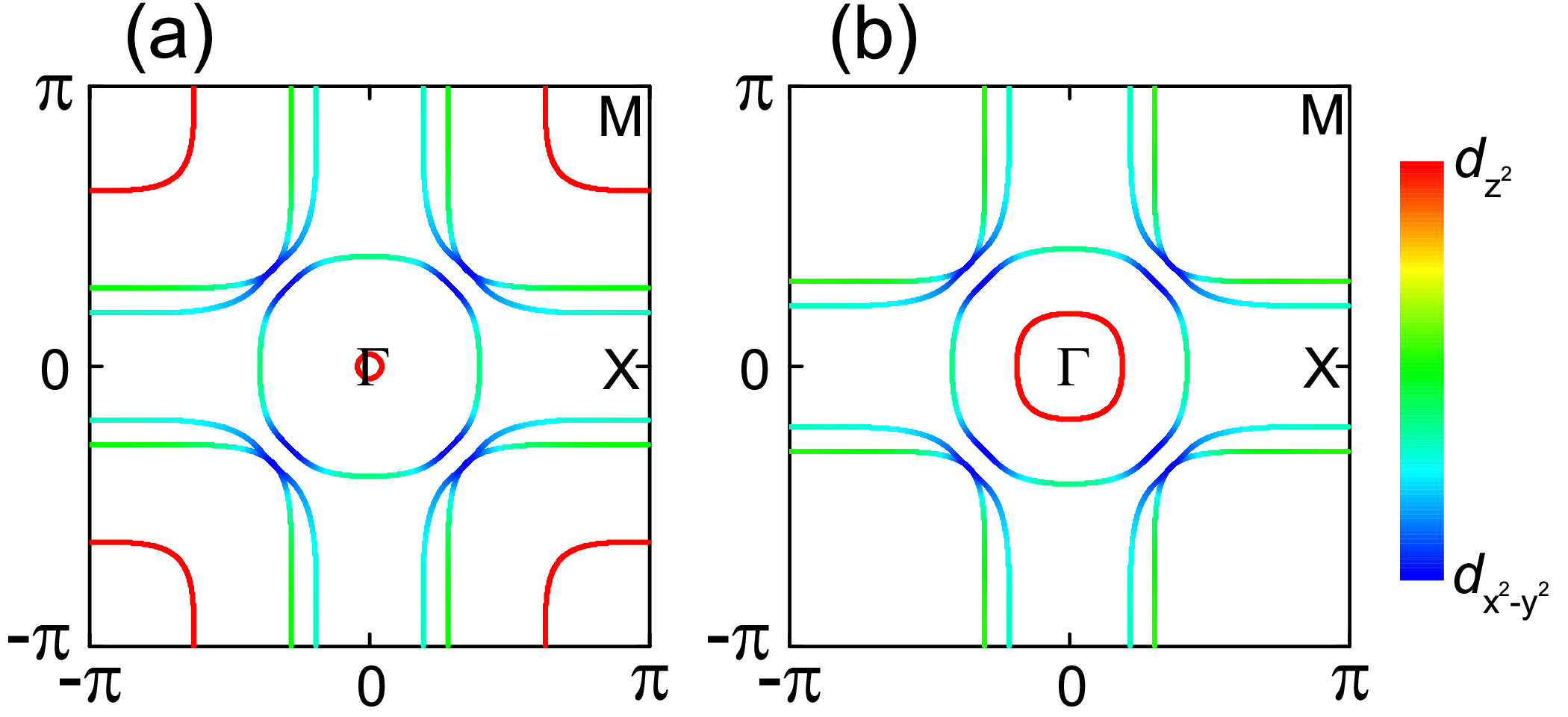}
\caption{FSs of $I4/mmm$ structure from effective two-orbital model under increasing electron doping. From (a) to (b), the doping concentration increases from 0.0015 to 0.03.
}
\label{FS_doping}
\end{figure}

In order to investigate the consequence of trilayer Ni-O planes on the electronic structure and SC pairing mechanism, we intend to extract an effective lattice model of the SC $I4/mmm$ phase of La$_4$Ni$_3$O$_{10}$. Firstly, we project the DFT bands into a subspace containing six Ni-$e_g$ orbits by using maximum-localized Wannier function implemented by Wannier90 package, thus obtain the tight-binding (TB) coefficients between these $e_g$ orbits. The hopping parameters in La$_4$Ni$_3$O$_{10}$ seem slightly greater than in La$_3$Ni$_2$O$_{7}$~\cite{Luo23} overall, which may imply that the former are less correlated. The corresponding Wannier bands are shown as dashed lines in Fig. \ref{bands1}(a), which show good agreement with the DFT bands. Afterwards, for the Wannier coefficients, we retain the TB coefficients between $e_g$ orbits to the order of next-next-nearest-neighbor, thus obtaining a simplified TB model on a two-dimensional square lattice describing the trilayer Ni-O planes, of which the model parameters are given explicitly in Tab. \ref{tab2} in the appendix. The resulting non-interacting electronic bands of the extracted simplified TB model are shown in Fig. \ref{bands1}(b) and (d), which can properly reproduce the orbital-projected and layer-projected representation of DFT bands in (a) and (c). The FSs calculated by the effective TB model are shown in Fig. \ref{FS}(a), indicating good agreement with the DFT FSs in Fig. \ref{FS-wannier}(b). Similar to the bilayer nikelate superconductor La$_3$Ni$_2$O$_{7}$~\cite{Sun23,Luo23}, La$_4$Ni$_3$O$_{10}$ also exhibits an electron pocket $\alpha$ around the $\Gamma$ point (contributed by two $e_g$ orbits, with average filling 0.12) and hole pockets $\beta$ around $M$ points (formed by $3d_{z^2}$ orbit, with average filling 0.81). In addition, La$_4$Ni$_3$O$_{10}$ possesses two nearby hole pockets $\gamma$ around $M$ point (constructed by mixture of two $e_g$ orbits), one of which is induced by the inner Ni-O plane (see Fig. \ref{FS_layer}), and the consequence of such doublet hole pockets to the superconductivity deserves deeper understanding. The large electron filling of $\beta$ hole pocket indicates that the $3d_{z^2}$ orbits are subjected to Coulomb repulsion, which should be correctly included in the effective TB model, see appendix. Due to proximity of the fifth band to the Fermi level (see Fig. \ref{bands1}(a)), a slight electron doping can induce a small electron pocket around $\Gamma$ point, see Fig. \ref{FS_doping}, the consequence of which to superconductivity should be further examined.

For $Cmca$ and $P2_1/a$ phases, the FSs on  $k_z=0$ plane can also be reproduced qualitatively by the effective TB model of $I4/mmm$ phase through Brillouin-zone folding along the boundaries of the extracted Brillouin zone (dashed lines), the result is illustrated in Fig. \ref{FS}(b), which agrees well with DFT FSs in Fig. \ref{FS-wannier} (c) and (d). It is worth noting that the SC phases in both La$_3$Ni$_2$O$_{7}$ and La$_4$Ni$_3$O$_{10}$ favor the aligned NiO$_6$ octahedrons perpendicular to the Ni-O planes~\cite{Sun23,Li23}, thus the physical origin of the disadvantage of tilted NiO$_6$ octahedrons to superconductivity should be further studied and understood.

\section{layer-dependent electronic correlation}
\begin{figure}[tbp]
\hspace{-0cm} \includegraphics[totalheight=2.2in]{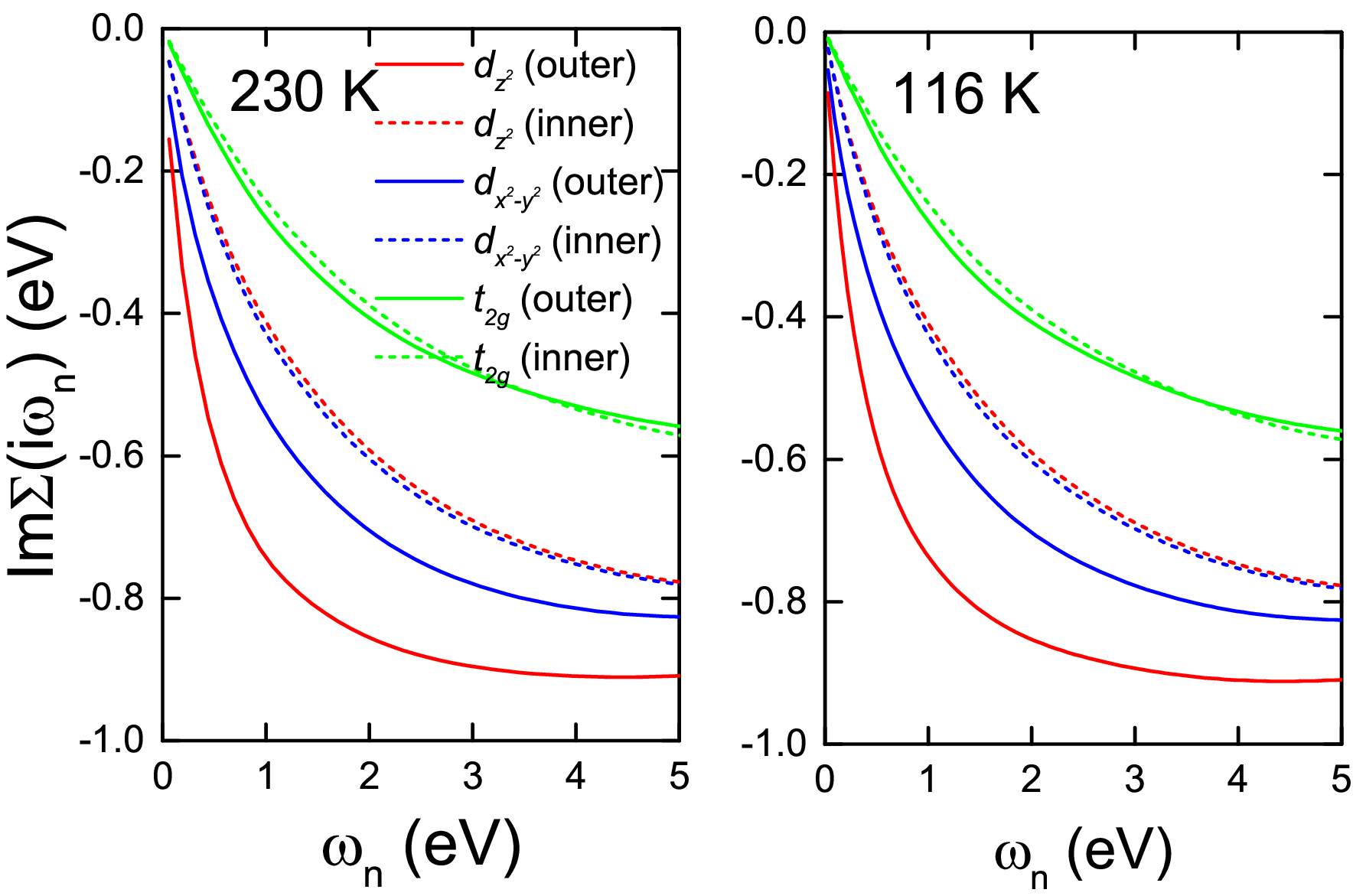}
\caption{DFT+DMFT self-energies of Ni atoms in $I4/mmm$ phase, on (a) imaginary- and (b) real-frequency axes, respectively.
}
\label{Sigma}
\end{figure}

\begin{figure*}[tbp]
\hspace{-0cm} \includegraphics[totalheight=2in]{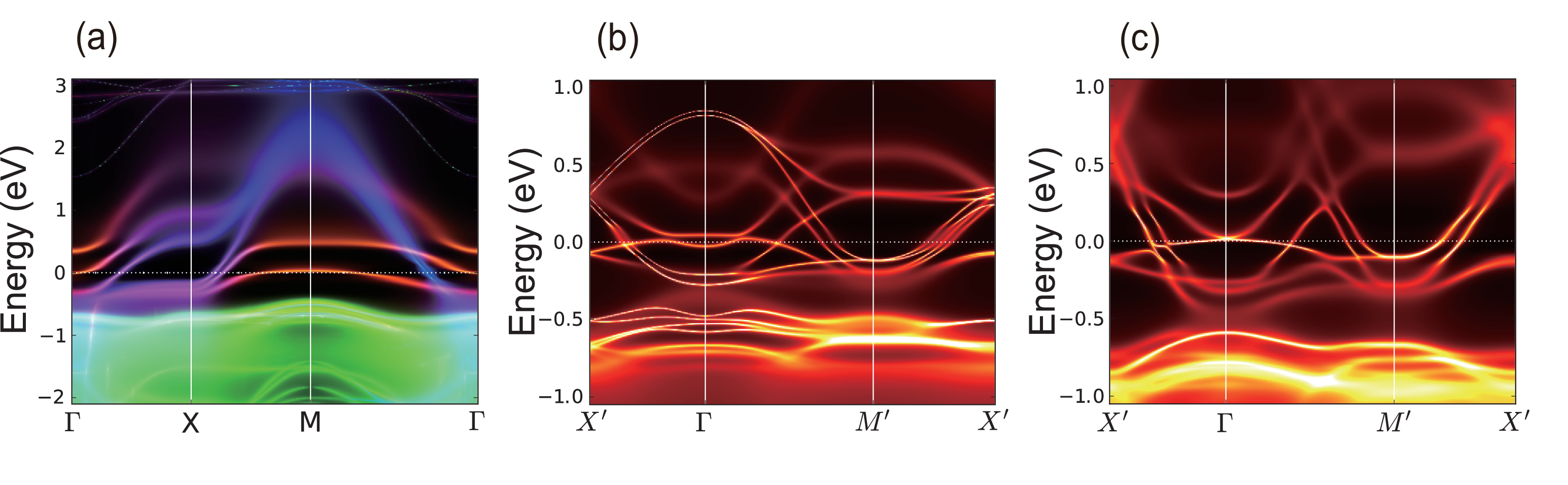}
\caption{Momentum-resolved spectral functions for (a) $I4/mmm$, (b) $Cmca$ and (c) $P2_1/a$ phases at 230 K, respectively. In (a) the red, blue and green colors denote the components of  $3d_{z^2}$, $3d_{x^2-y^2}$ and $t_{2g}$ orbits, respectively.
}
\label{specfunc}
\end{figure*}

Since the superconductivity of La$_4$Ni$_3$O$_{10}$ emerges after the density-wave like magnetic order is suppressed at high pressure~\cite{Zhang2311,Zhu23}, the electronic correlations particularly the spin correlations may play a crucial role in the SC pairing mechanism. In order to study the electronic correlations, we employ the DFT+DMFT algorithm embodied in EDMFT package~\cite{Haule10}, which is a powerful tool in simulating the correlated materials. We employ the on-site Coulomb
repulsion $U$=5.0 eV and Hund's coupling $J_H$=0.7 eV on
Ni-3d orbits, which are typical value used for nickelates~\cite{Ouyang23}, and we use the hybridization expansion version of the continuous-time quantum Monte Carlo method (CT-QMC) as the impurity solver. Fig. \ref{Sigma} shows the imaginary parts of self-energies of Ni atoms, which indicates a notable orbital and layer dependence. In Fig. \ref{specfunc}(a), we illustrate the momentum-resolved spectral function of $I4/mmm$ phase at temperature $T$=230 K, in which the spectral weights of two $e_g$ orbits are represented by different colors. Near the Fermi level, the momentum-resolved spectral functions turn out to be close related to the DFT bands in Fig. \ref{bands1}(a). The similarity between DFT and DFT+DMFT bands suggest that the electronic correlations give only minor modification to the FSs, therefore the FSs deduced by the non-interacting part of the effective TB model in Fig. \ref{FS}(a) can provide appropriate description to the FSs even at presence of electronic correlations~\cite{Leonov24}. Noticeably, the energy spreading of DFT+DMFT bands are remarkably reduced in comparison to the DFT bands, leading to enhancement of quasi-particle mass $m^*/m_{DFT}$, which can be evaluated through the quasi-particle spectral weight $Z=1/(1-\frac{\partial\Sigma(\omega_n)}{\partial\omega_n}|_{\omega_n\rightarrow0^+})$ by $m^{\ast}/m_{DFT}=1/Z$, with the results listed in Tab. \ref{tab1}). Ni-$e_g$ orbits on the outer layers have considerably greater mass enhancements than on the inner layer, which reflects that the outer layers are more correlated and may be more responsible to the formation of superconductivity. Beside, on outer layers, $d_{z^2}$ orbit has a much greater mass enhancement than $d_{x^2-y^2}$ and $t_{2g}$ orbits, implying $d_{z^2}$ orbits are more correlated, in accordance with the flat $3d_{z^2}$ bands in the spectral function in Fig. \ref{specfunc}(a). The strongly-correlated electronic character indicates an unconventional SC pairing mechanism, therefore, the electronic correlations should be treated properly, particularly for the correlated $3d_{z^2}$ orbits which has a filling factor $\sim0.81$ remarkably higher than half filling. It should be noted that the mass enhancements on the outer (inner) Ni-O layers in La$_4$Ni$_3$O$_{10}$ are stronger (weaker) than trilayer La$_3$Ni$_2$O$_{7}$ (which is 2.74 at 220 K~\cite{Ouyang23}), indicating that in La$_4$Ni$_3$O$_{10}$ the SC pairing mechanism is rather distinct to La$_3$Ni$_2$O$_{7}$, and should be deeply investigated in the context of the trilayer structure.

The spectral functions of $Cmca$ and $P2_1/a$ phases are illustrated in Fig. \ref{specfunc}{(b-c). Beside the reduction of energy spreadings, the spectral functions are compatible to the DFT bands in Fig. \ref{bands2}(a) and (b), respectively, which in turn confirms the validity of Brillouin-zone folding in the understanding of DFT bands and FSs of $Cmca$ and $P2_1/a$ phases in terms of $I4/mmm$ bands, as previously shown in Fig. \ref{bands2}(c) and Fig. \ref{FS}(b).

\heavyrulewidth=1bp

\begin{table}
\small
\renewcommand\arraystretch{1.3}
\caption{\label{tab1}
Quasi-particle mass enhancements $m^{\ast}/m_{DFT}$ of Ni-$3d$ states in $I4/mmm$ phase of La$_4$Ni$_3$O$_{10}$ through DFT+DMFT calculation, which show slight increase upon cooling.}
\begin{tabular*}{8cm}{@{\extracolsep{\fill}}cccc}
\toprule
    &    & 116 K & 230 K  \\
\hline
 &$d_{z^2}$   & 3.72  &        3.47                 \\
 outer layers &$d_{x^2-y^2}$   & 2.69  &        2.52                            \\
 &$t_{2g}$   &  1.32 &        1.32 \\
  \hline
 & $d_{z^2}$   &  1.75 &        1.73                 \\
 inner layer & $d_{x^2-y^2}$   &  1.73 &        1.74                            \\
 &$t_{2g}$   & 1.28  &        1.28                              \\
\bottomrule
\end{tabular*}
\label{tab1}
\end{table}

\section{conclusion and discussion}

To summarize, we have explored the electronic structures and electronic correlations in distinct phases of La$_4$Ni$_3$O$_{10}$ under different pressure regions. We suggest that under ambient and low pressures, the electronic bands and Fermi surfaces of $Cmca$ and $P2_1/a$ phases can be constructed qualitatively through the superconducting $I4/mmm$ phase in the context of Brillouin-zone folding, as a consequence of tilted NiO$_6$ octahedrons in $Cmca$ and $P2_1/a$ phases. Similar to the trilayer nickelate superconductor La$_3$Ni$_2$O$_{7}$, in SC $I4/mmm$ phase of La$_4$Ni$_3$O$_{10}$, the electronic bands near the Fermi level are dominated by the $e_g$ orbits of Ni-$3d$ states, in which the $3d_{x^2-y^2}$ orbit constructs a electron pocket around the Brillouin center, and the $3d_{z^2}$ state forms hole pockets around Brillouin-zone corners. Notably, the additional inner Ni-O plane in SC La$_4$Ni$_3$O$_{10}$ gives rise to a doublet hole pockets surrounding Brillouin zone corners, which are contributed by mixture of $3d_{x^2-y^2}$ and $3d_{z^2}$ states. Besides, the presence of inner Ni-O plane pushes one branch of $3d_{z^2}$ bands much closer to the Fermi level, thus can induce a small $3d_{z^2}$ electron pocket around the Brillouin-zone center under slight electron doping. By using DFT+DMFT simulations of the SC $I4/mmm$ phase, we verify that the electronic states near the Fermi level are dominated by Ni-3$d_{x^2-y^2}$ and 3$d_{z^2}$ states, though the Fermi surface structure are only weakly modified by electronic correlations, the electronic states near the Fermi level are correlated with remarkably enhanced quasi-particle mass, particularly for the Ni-3$d_{z^2}$ state. Notably, $3d$ states on the outer layers are considerably more correlated than on the inner layer.
In this context, we have extracted an effective two-orbital tight-binding model for the $I4/mmm$ phase, which is appropriate for further studying the SC pairing mechanism and pairing symmetry in trilayer nickelate La$_4$Ni$_3$O$_{10}$.

In comparison to the bilayer La$_3$Ni$_2$O$_{7}$, the hopping strengths in La$_4$Ni$_3$O$_{10}$ are overall greater, leading to weaker electron correlations than the former.
Since the inter-layer vertical hybridization between $d_{z^2}$ orbits ($e_d=-0.7$ eV) dominates over the intra-layer hybridization ($t_d=-0.15\sim-0.16$ eV), the leading spin correlation is the inter-layer spin-singlet pairing $\chi^\perp_s$, which is induced by the superexchange coupling between inter-layer $d_{z^2}$ bonds with strength $J^\perp_d=e^2_d/U$~\cite{YFYang23}. Under the trilayer structure, the inter-layer spin-singlet pairing in La$_4$Ni$_3$O$_{10}$ may be partially reduced with respect to the bilayer La$_3$Ni$_2$O$_{7}$, due to the inner layer which interacts with both outer layers, causing some frustration-like effects. The possible reduction of spin-singlet pairing may further give rise to the consequence that the SC pairing in La$_4$Ni$_3$O$_{10}$ may be dominated by two outer layers which have greater electronic and spin correlations, and additional, the inter-layer hybridization of $d_{z^2}$ orbits between two outer layers ($g_d=-0.105$ eV) leading to an unignorable superexchange $J=g^2_d/U$ which acts as a frustration and further reduce the spin-singlet pairing $\chi^\perp_s$. In this context, the lower SC transition temperature in La$_4$Ni$_3$O$_{10}$ than La$_3$Ni$_2$O$_{7}$ may attribute to weaker correlation effect and the trilayer structure, which can be examined in further study based on our effective TB model. In addition, the FSs of $d_{z^2}$ state in La$_4$Ni$_3$O$_{10}$ are sensitive to electron doping (Fig. \ref{FS_doping}), which may lead to drastic influence to superconductivity.

\section{Appendix: Description of the effective tight-binding model}

For the trilayer Ni-O planes, by defining
$
\mathbf{\Phi}_{\mathbf{k}\sigma}=(c^{(1)}_{\mathbf{k}\sigma}d^{(1)}_{\mathbf{k}\sigma}c^{(2)}_{\mathbf{k}\sigma}d^{(2)}_{\mathbf{k}\sigma}c^{(3)}_{\mathbf{k}\sigma}d^{(3)}_{\mathbf{k}\sigma})
$, in which the superscripts 1, 2, 3 denote the upper, inner and lower Ni-O planes, $c$ and $d$ represent the annihilate operators for Ni-$3d_{x^2-y^2}$ and $3d_{z^2}$ orbits, respectively, the non-interacting part of the effective TB Hamiltonian can be written as a compact matrix form
$
\mathcal{H}_0=\sum_{\mathbf{k}\sigma}{\bf\Phi}^{\dag}_{\mathbf{k}\sigma}\mathbf H_{\mathbf k}\bf\Phi_{\mathbf{k}\sigma}
$, in which the summation of $\mathbf{k}$ is restricted in the Brillouin zone of square lattice, and the Hamiltonian matrix reads
\begin{align}
\mathbf{H}_{\mathbf{k}}=
\left(
\begin{array}{ccc}
\mathbf{H}_{11} & \mathbf{H}_{12} & \mathbf{H}_{13}\\
\mathbf{H}_{21}& \mathbf{H}_{22}& \mathbf{H}_{23}\\
\mathbf{H}_{31}& \mathbf{H}_{32}& \mathbf{H}_{33}
\end{array}
\right)=\left(
\begin{array}{ccc}
\mathbf{H}_{11} & \mathbf{H}_{12} & \mathbf{H}_{13}\\
\mathbf{H}_{12}& \mathbf{H}_{22}& \mathbf{H}_{12}\\
\mathbf{H}_{13}& \mathbf{H}_{12}& \mathbf{H}_{11}
\end{array}
\right)
\label{Hk}.\end{align}
Since the two outer planes are equal to each other, the above equation has taken the relations $\mathbf{H}_{33}=\mathbf{H}_{11}$, $\mathbf{H}_{23}=\mathbf{H}_{12}$, with
\begin{equation*}
\mathbf{H}_{11}=
\left(
\begin{array}{cc}
\lambda_\mathbf{k}(\epsilon^{(o)}_c,t^{(o)}_c,t^{\prime (o)}_c,t^{\prime\prime (o)}_c) & \gamma_\mathbf{k}(t^{(o)}_{dc},t^{\prime\prime (o)}_{dc})  \\
\gamma_\mathbf{k}(t^{(o)}_{dc},t^{\prime\prime (o)}_{dc}) & \lambda_\mathbf{k}(\epsilon^{(o)}_d,t^{(o)}_d,t^{\prime (o)}_d,t^{\prime\prime (o)}_d)
\end{array}
\right)
,\end{equation*}
\begin{equation*}
\mathbf{H}_{22}=
\left(
\begin{array}{cc}
\lambda_\mathbf{k}(\epsilon^{(i)}_c,t^{(i)}_c,t^{\prime (i)}_c,t^{\prime\prime (i)}_c) & \gamma_\mathbf{k}(t^{(i)}_{dc},t^{\prime\prime (i)}_{dc})  \\
\gamma_\mathbf{k}(t^{(i)}_{dc},t^{\prime\prime (i)}_{dc}) & \lambda_\mathbf{k}(\epsilon^{(i)}_d,t^{(i)}_d,t^{\prime (i)}_d,t^{\prime\prime (i)}_d)
\end{array}
\right)
,\end{equation*}
\begin{equation*}
\mathbf{H}_{12}=
\left(
\begin{array}{cc}
\lambda_\mathbf{k}(e_c,s_c,s^{\prime }_c,s^{\prime\prime }_c) & \gamma_\mathbf{k}(s_{dc},s^{\prime\prime }_{dc})  \\
\gamma_\mathbf{k}(s_{dc},s^{\prime\prime }_{dc}) & \lambda_\mathbf{k}(e_d,s_d,s^{\prime }_d,s^{\prime\prime }_d)
\end{array}
\right)
,\end{equation*}
\begin{equation*}
\mathbf{H}_{13}=
\left(
\begin{array}{cc}
\lambda_\mathbf{k}(g_c,r_c,r^{\prime }_c,r^{\prime\prime }_c) & \gamma_\mathbf{k}(r_{dc},r^{\prime\prime }_{dc})  \\
\gamma_\mathbf{k}(r_{dc},r^{\prime\prime }_{dc}) & \lambda_\mathbf{k}(g_d,r_d,r^{\prime }_d,r^{\prime\prime }_d)
\end{array}
\right)
,\end{equation*}

in which $\lambda_\mathbf{k}(d,t,t^\prime,t^{\prime\prime}) =
d+2t(\cos k_x+\cos k_y)+4t^\prime\cos k_x\cos k_y+2t^{\prime\prime}(\cos 2k_x+\cos 2k_y)$, $\gamma_\mathbf{k}(t,t^{\prime\prime})=
2t(\cos k_x-\cos k_y)+2t^{\prime\prime}(\cos 2k_x-\cos 2k_y)$. As shown in Fig. \ref{hopping}, in the effective TB model, the hopping parameters extracted from the Wannier coefficients are approximated to the order of next-next-nearest-neighbor (with displacement $2\mathbf{a}$ or $2\mathbf{b}$), with strength (in unit of eV) listed in Tab. \ref{tab2}. The chemical potential $\mu=10.83$ eV is set equal to the Fermi energy in DFT. In these parameters, $\epsilon$ denotes local potential; $e$ and $g$ denote hybridization among neighbor layers and two outer layers along perpendicular direction, respectively; $t$ denotes intra-layer hopping strength; $s$ denotes the hybridization between nearby layers; $r$ denotes the hybridization between two outer layers. The subscripts $dc$ denote the hybridization between $d$ and $c$ orbits. During the Wannier fitting, the TB parameters are found to be slightly different between inner and outer layers, for which the parameters are denoted by superscripts $(i)$ and $(o)$, respectively. The nearest-neighbor, next-nearest-neighbor and next-next-nearest-neighbor (measuring in terms of on-plane projection) components in $t$, $s$ and $r$ are denoted by no superscript and superscripts $'$ and $''$, respectively, see Fig. \ref{hopping}.
\begin{figure}[tbp]
\hspace{-0cm} \includegraphics[totalheight=2.5in]{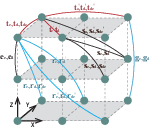}
\caption{Tight-binding parameters in the effective TB models on trilayer Ni-O planes. The green dots denote Ni atoms, among which the red, black and blue links denote the intra-plane, inner-outer and outer-outer hopping parameters. Corresponding hopping strengths are listed in Tab. \ref{tab2}.
}
\label{hopping}
\end{figure}
Since the Ni-$3d_{z^2}$ state constructs $\beta$ hole pockets with filling factor 0.81 remarkably larger than half-filling, the on-site Coulomb repulsion between $3d_{z^2}$ electrons should be included in the effective TB model, with the interacting part as $
\mathcal{H}_{int}=\sum_{i,l}Un^{(l)}_{di\uparrow}n^{(l)}_{di\downarrow}
$, where $l=1,2,3$ denote layer numbers, then the entire Hamiltonian of the effective TB model reads $\mathcal{H}=\mathcal{H}_{0}+\mathcal{H}_{int}$.

\heavyrulewidth=1bp

\begin{table*}
\small
\renewcommand\arraystretch{1.3}
\caption{\label{tab2}
Hopping parameters (illustrated in Fig. \ref{hopping}) in $I4/mmm$ phase of La$_4$Ni$_3$O$_{10}$, in units of eV. $c$ and $d$ denote $d_{x^2-y^2}$ and $d_{z^2}$ orbits. Some hybridizations are diminished due to the orbital symmetry, indicated by "-".}
\begin{tabular*}{18cm}{@{\extracolsep{\fill}}cccccc}
\toprule
              &    hybridization      & vertical & 1st neighbor ($\mathbf{a}$) & 2nd neighbor ($\mathbf{a}+\mathbf{b}$) &  3rd neighbor ($2\mathbf{a}$) \\
\hline
              & $c-c$  & $\epsilon^{(i)}_c=11.98$ & $t^{(i)}_c=-0.511$  & $t^{\prime (i)}_c=0.051$  & $t^{\prime\prime (i)}_c=-0.082$ \\
intra-plane (inner)   & $d-d$  & $\epsilon^{(i)}_d=11.88$ & $t^{(i)}_d=-0.162$  & $t^{\prime (i)}_d=-0.017$  & $t^{\prime\prime (i)}_d=-0.026$ \\
              & $c-d$  & - & $t^{(i)}_{dc}=0.296$  & - & $t^{\prime\prime (i)}_{dc}=0.041$ \\
  \hline
              & $c-c$  & $\epsilon^{(o)}_c=11.63$ & $t^{(i)}_c=-0.516$  & $t^{\prime (o)}_c=0.064$  & $t^{\prime\prime (o)}_c=-0.067$ \\
intra-plane (outer)   & $d-d$  & $\epsilon^{(o)}_d=11.49$ & $t^{(o)}_d=-0.151$  & $t^{\prime (o)}_d=-0.014$  & $t^{\prime\prime (o)}_d=-0.021$ \\
              & $c-d$  & - & $t^{(o)}_{dc}=0.283$  & - & $t^{\prime\prime (o)}_{dc}=0.036$ \\
  \hline
              & $c-c$  & $e_c=0.011$ & $s_c=0.002$  & $s^\prime_c=0.004$  & $s^{\prime\prime}_c=-0.005$ \\
inner-outer   & $d-d$  & $e_d=-0.7$ & $s_d=0.038$  & $s^\prime_d=0.004$  & $s^{\prime\prime}_d=-0.001$ \\
              & $c-d$  & - & $s_{dc}=-0.037$  & -  & $s^{\prime\prime}_{dc}=0.003$ \\
  \hline
              & $c-c$  & $g_c=-0.0026$ & $r_c=0$  & $r^\prime_c=-0.0013$  & $r^{\prime\prime}_c=0$ \\
outer-outer   & $d-d$  & $g_d=-0.105$ & $r_d=-0.01$  & $r^\prime_d=0.003$  & $r^{\prime\prime}_d=0.0018$ \\
              & $c-d$  & - & $r_{dc}=-0.0015$  & - & $r^{\prime\prime}_{dc}=0$ \\
\bottomrule
\end{tabular*}
\label{tab2}
\end{table*}

\acknowledgments
This work is supported by National Natural Science Foundation of China (No. 12364023),  and GuikeAD20159009.

\end{document}